\documentclass{elsart}
\usepackage{epsf}

\begin{document}
\begin{frontmatter}

\title{Time scales involved in market emergence}

\author{J. Kwapie\'n$^{1,2}$, S. Dro\.zd\.z$^{1-3}$ and J. Speth$^1$}
\address {$^1$Institut f\"ur Kernphysik, Forschungszentrum J\"ulich,
D-52425 J\"ulich, Germany \\
$^2$Institute of Nuclear Physics, PL--31-342 Krak\'ow, Poland \\
$^3$Institute of Physics, University of Rzesz\'ow, PL--35-310 Rzesz\'ow,
Poland}

\begin{abstract}

In addressing the question of the time scales characteristic for the
market formation, we analyze high frequency tick-by-tick data from the
NYSE and from the German market. By using returns on various time scales
ranging from seconds or minutes up to two days, we compare magnitude of
the largest eigenvalue of the correlation matrix for the same set of
securities but for different time scales. For various sets of stocks of
different capitalization (and the average trading frequency), we observe a
significant elevation of the largest eigenvalue with increasing time
scale. Our results from the correlation matrix study go in parallel with
the so-called Epps effect. There is no unique explanation of this effect
and it seems that many different factors play a role here. One of such
factors is randomness in transaction moments for different stocks. Another
interesting conclusion to be drawn from our results is that in the
contemporary markets the emergence of significant correlations occurs on
time scales much smaller than in the more distant history. 

\end{abstract}

\end{frontmatter}

\section{Introduction}

A number of studies carried out over the past years showed that the time
evolution of individual securities depends strongly on the evolution of
other securities and even of the whole market. This observation leaded to
the development of various models which help investors to minimize risk
and to choose the optimal investment strategies. Therefore the magnitude,
the temporal stability and the time-scale characteristics of the
correlations are crucial factors for these models. Also from the
theoretical point of view the existence and strength of correlations play
an important role in development of proper models of the stock market
dynamics and could help understanding the mechanisms which are responsible
for the emergence of the collective signals out of noise in complex
systems.

It is well-documented in literature that what dominates the market
dynamics at the microscopic level is noise~\cite{bouchaud,madhavan,farmer}.
The movements of stock's price are governed by a series of buy/sell orders
reaching the market essentially at random moments (although there are some
long-time dependences which can be a source of the non-Gaussian tails of
the distributions of the inter-transaction time intervals). Among the
factors leading to this microscale randomness there is a difference in
reaction times of the investors to arriving news. This can be well related
to the different investment time horizons, acting through different market
makers and so on. Thus, even though an important piece of news arrives on
the market, different investors absorb this information and adjust their
positions at distinct moments. Similarly, there exists randomness in the
transaction volume, bid-ask spread etc. On short time scales, all these
elements cause the price to fluctuate stochastically around its ``true''
value like it is in a random walk. In these circumstances, if there is
some amount of persistence in the price evolution, it can be observed only
after many transactions take place, i.e. on longer time scales.

The above-described randomness in the price movements is even better
evident after a parallel inspection of the tick-by-tick data for two or
more assets is made. Apart from the-already-mentioned difference in the
reaction time of the investors for the same piece of news, there is a
separate news flow regarding each of the companies under study, which can
be yet another source of randomness. Therefore, we can safely assume that
on very short time scales comparable with the mean time interval between
consecutive trades, the correlations among the stock price fluctuations do
not differ from the noise level and they are insignificant. Consequently,
on such short time scales it is not justified to consider the market as a
coherent whole; one rather deals with a set of the elements evolving
independently from each other. Going from short time scales to the ones
much longer than the mean inter-transaction interval, new effects occur. 
Firstly, all the investors have opportunity to react to the news, which
gives the complete picture of how this piece of news affects the price. 
Secondly, the investors analyse price changes of other assets and correct
their positions and strategies accordingly. In the price evolution of each
asset there is thus information of other assets' prices which causes the
inter-stock correlations to emerge. This diffusion of information
increases with increasing time, bringing about the correlations to be
strenghtened either. Moreover, the longer time passed, the more investors
manage to act, which leads to an even wider flow of information between
the assets and the time scales~\cite{mueller}. From this angle, the
coupling strength reaches its maximum after all the investors can correct
their positions. Macroscopically, the existence of the inter-stock
correlations, both those originating from similar responses of different
stocks to the same piece of news and those being the effect of a directed
network of influence among the stocks~\cite{kullmann}, is an important
aspect of the collective market formation (see also~\cite{bonanno}). The
most striking evidence of this phenomenon is the strong index movements
and trends which cannot be observed in a completely decorrelated system.

In our paper, we would like to address the closely related question of
what are the time scales at which the significant correlations emerge. 
i.e. at which time scale range the transition from noise to a collective
behaviour takes place, and what are the factors responsible for the
inter-stock coupling strength. We shall compare the properties of the
inter-stock correlations at several different time scales for different
groups of stocks by applying the globally-oriented correlation matrix
formalism in order to look at the couplings between more than two assets. 
Additionally, motivated by the results from our earlier
study~\cite{drozdz03b} dealing with the statistical properties of the
stock price fluctuations, we intend to compare the correlation properties
of the contemporaneous and the historical data. 

\section{Results}

We analyze high frequency data from the American and from the German stock
market contained in the TAQ and the KKMDB databases~\cite{data},
respectively. In both cases, the data covers the over-two-years-long
period from Dec 1, 1997 to Dec 31, 1999.  For each company listed on NYSE
and NASDAQ markets, the TAQ database contains a record of all transactions
which took place within a given time interval of trading (the same refers
to the KKMDB database and the Deutsche B\"orse). As the transactions are
made at random moments, first we need to create a time series of price
values being sampled with constant frequency. Following the standard
prescription, we assume that the price $x_{\alpha}(t_i)$ of an asset
$\alpha$ at time $t_i$ is equal to the price of the last preceiding
transaction on the corresponding stock.  Then, given a time scale $\Delta
t$, we calculate a time series of normalized returns defined by
\begin{equation} 
g_{\beta}(t_i)=\frac {G_{\beta}(t_i) - \langle
G_{\beta}(t_i) \rangle_{t_i}} {\sigma(G_{\beta})}, \ \quad
\sigma(G_{\beta}) = \sqrt{\langle G_{\beta}^2(t_i) \rangle_{t_i} - \langle
G_{\beta}(t_i) \rangle_{t_i}^2} \end{equation} where \begin{equation}
G_{\beta}(t_i) = \ln x_{\beta}(t_i+\Delta t) - \ln x_{\beta}(t_i). 
\end{equation} 
Here $\langle \ldots \rangle_{t_i}$ stands for averaging over discrete
time. Let us say we have a set of $N$ stocks and from the corresponding
time series of length $T$ we construct an $N \times T$ matrix ${\bf M}$
and finally we calculate an $N \times N$ correlation matrix ${\bf C}$
according to the formula
\begin{equation} 
{\bf C} = (1 / T) \ {\bf M} {\bf M}^{\rm T}.
\end{equation} 
Each element $C_{\alpha,\beta}$ of the correlation matrix is simply the
correlation coefficient for the pair of stocks $\alpha$ and $\beta$. 
Afterwards, the correlation matrix can be diagonalized in order to obtain
the spectrum of its eigenvalues $\lambda_k$
($k=1,...,N$)~\cite{laloux,plerou99a,drozdz00,drozdz01,kwapien02}.  We
repeat this procedure for several distinct time scales ranging from 1 min
(or even 1 s) up to two trading days (780 min in New York and 1020 min in
Frankfurt). Typically for the stock market data, the correlation matrix
develops at least one eigenvalue which is repelled from the rest of the
spectrum and which corresponds to a collective behaviour of group of
stocks or a whole
market~\cite{laloux,plerou99a,drozdz00,drozdz01,kwapien03}. It is
convenient to confront the eigenvalue spectrum with the universal
predictions of the Random Matrix Theory as all the apparent discrepancies
can be related to the existence of market-specific information.

We start from an investigation of the coupling strength's dependence on
the time scale $\Delta t$ for pairs of stocks. It is convenient to
quantify this dependence in terms of the correlation coefficient
$C_{\alpha,\beta}(\Delta t)$. In Figure 1 we show this quantity for two
different but typical pairs of the DJI stocks: Alcoa (AA) $-$ Exxon (XON)
and Chevron (CHV) $-$ Exxon. This Figure shows that $C_{\alpha,\beta}$ is
definitely not invariant under a change of the time scale. Essentially,
the magnitude of the correlation coefficient increases while going from
the smaller intra-hour or minute to the larger daily $\Delta t$'s. As CHV
and XON belong to the same market sector (energy), while AA does not, for
each $\Delta t$ the correlations in the former case are much more
significant then in the latter one. Nevertheless, both plots demonstrate
only small correlations at the shortest analyzed time scale of 1 min while
for larger $\Delta t$ the correlation coefficient significantly increases.
For both pairs of stocks the picture is qualitatively similar up to
$\Delta t\simeq 20$ min but then we observe a difference in the behaviour
of the correlations for larger $\Delta t$'s: $C_{\rm CHV,XON}$ still goes
up with increasing the time scale reaching as high as 0.65 for daily
returns, while $C_{\rm AA,XON}$ almost saturates below 0.20. However, a
trace of saturation is also identifiable for the CHV-XON pair.

\begin{figure}
\epsfxsize 11cm
\hspace{1.0cm}
\epsffile{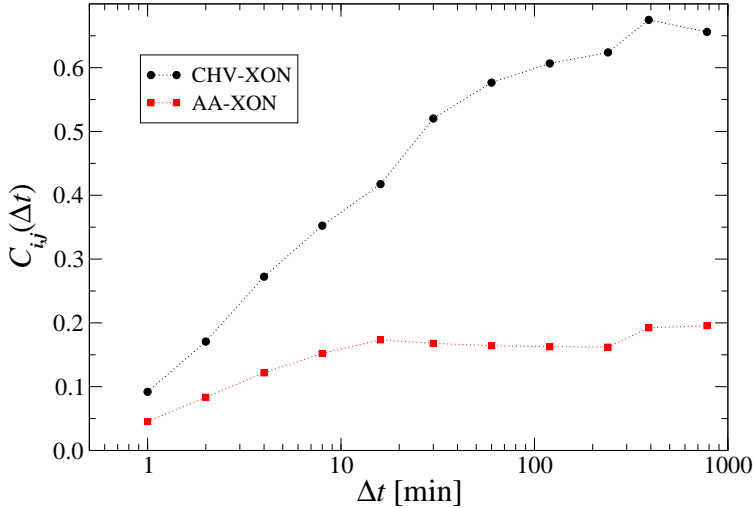}
\caption{Correlation coefficient $C_{ij}$ as a function of time scale
$\Delta t$ for two exemplary pairs of NYSE companies representing the same
market sector (CHV and XON, circles) or two different sectors (AA and CHV,
squares).}
\label{fig:effect}
\end{figure}

Due to the fact that such an increase of correlation magnitude is
characteristic for all pairs of stocks, one may expect that a similar
effect can also be observed if one looks at a more global measure of
correlations e.g. the largest eigenvalue of the correlation matrix. Thus
we select two sets of 30 stocks listed in the DJIA and DAX indices and we
evaluate the corresponding function $\lambda_1(\Delta t)$; the results are
displayed in Figure 2. Indeed, for both DJIA (Fig.~2(a)) and DAX
(Fig.~2(b)) this effect is strong, although the detailed behaviour of the
largest eigenvalue is not market invariant. In DJIA, $\lambda_1$ increases
for $\Delta t$ up to 30 min; for longer time scales there is no further
increase but rather a saturation of $\lambda_1$ can be seen. In contrast,
the largest eigenvalue for DAX gradually rises up to a daily time scale
($\Delta t=510$ min); this increase, however, is rather slow for $\Delta t
\ge 30$ min. These results show how the markets change their behaviour
from decorrelated and completely noisy dynamics to the collective one. 
This observation can be compared with earlier analysis of a market-sector
formation while going from short to long time scales presented in
ref.~\cite{bonanno}. Interestingly, for all time scales the DAX stocks
seem to be more strongly coupled than their DJI counterparts; for $\Delta
t=1$ min $\lambda_1$ for DJI only moderately differs from the RMT
prediction ($\lambda_1^{\rm RMT}(\Delta t=1 min) \simeq
1.0$)~\cite{sengupta}. The innately more correlated nature of the German
market, which leaves its fingerprints especially for $\Delta t > 30$ min,
has already been pointed out earlier (see eg.~\cite{drozdz01}) for daily
returns. Magnitude of this effect is substantially time-dependent, however
(compare with the results from 1990-2001 in~\cite{drozdz01}). 

These results go in parallel with the so-called Epps effect~\cite{epps},
named after the first researcher who demonstrated that when going from the
daily to the intra-hour time scales the inter-stock correlations decay.
Althoug his analysis was entirely based on the stock market data, the
similar results were obtained later for the currency exchange markets as
well (see~\cite{lundin,reno} for recent results). 

\begin{figure}
\epsfxsize 11cm
\hspace{1.0cm}
\epsffile{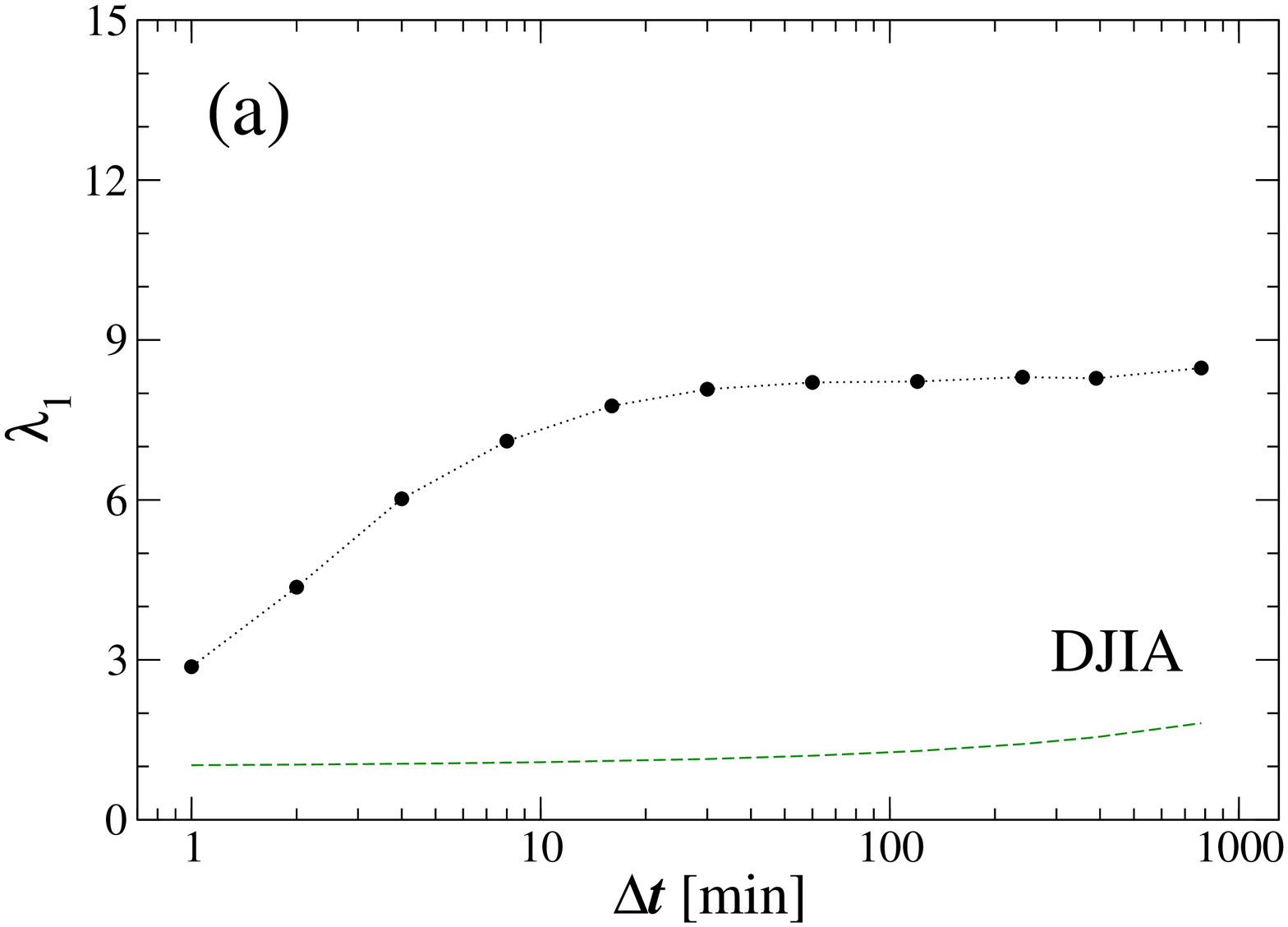}

\hspace{1.0cm}
\epsfxsize 11cm
\epsffile{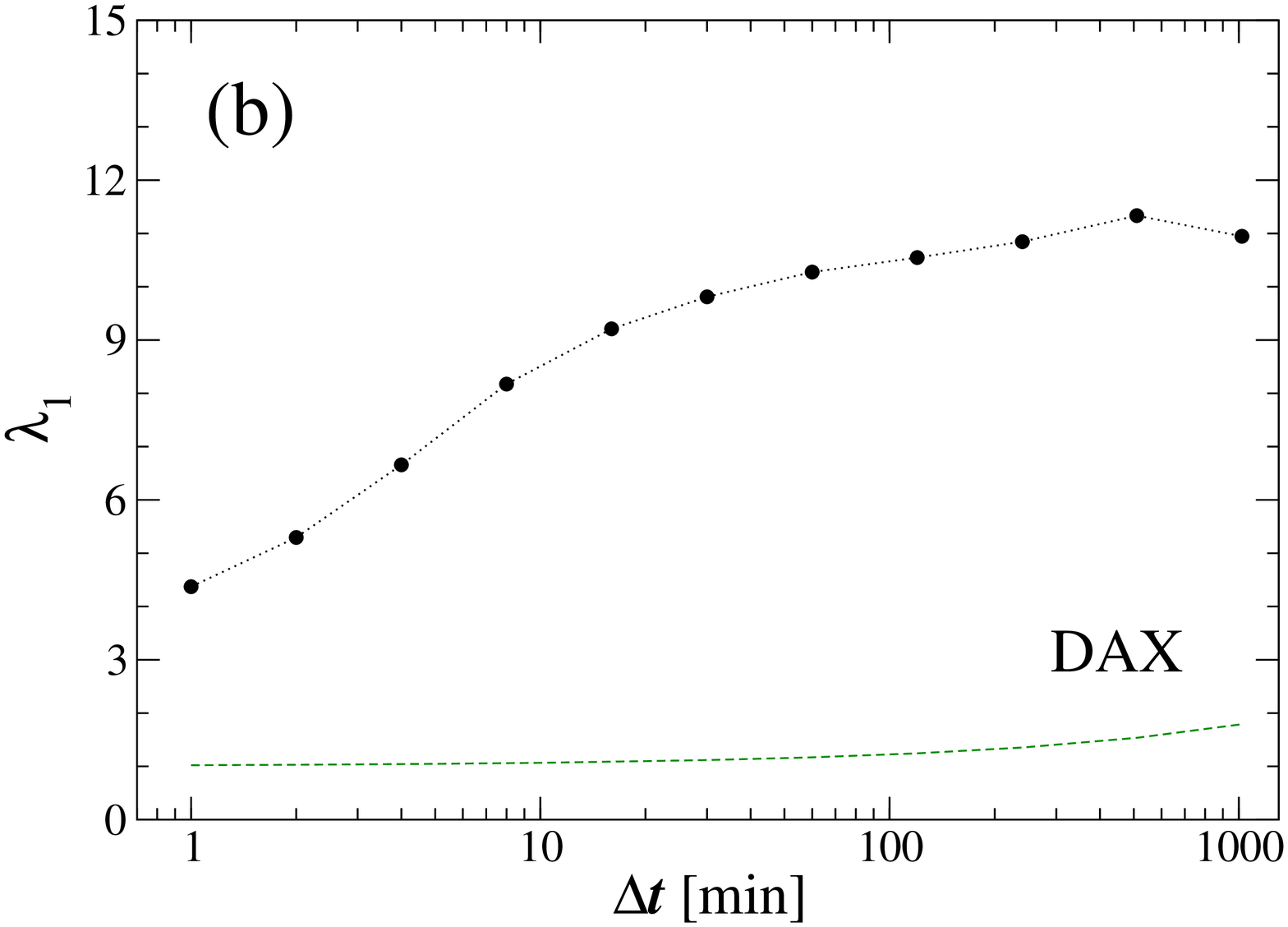}
\caption{Largest correlation matrix eigenvalue $\lambda_1$ as a function
of the time scale $\Delta t$ for 30 DJI stocks (a) and for 30 DAX stocks
(b); the Random Matrix Theory prediction for $\lambda_1^{\rm RMT}(\Delta
t)$, which depends on time series length~\cite{sengupta}, is denoted by
dashed lines.}
\label{fig:market}
\end{figure}

Price correlations among different stocks at high frequencies are caused
by market makers who quickly react to important news and to changes of
prices of other securities. As a security's price can be determined only
in a transaction, statistically a piece of news influences those
securities first which are traded more frequently than others. This of
course implies that also correlations are more likely to occur earlier for
the most active securities. This can be seen for the currency exchange
rates where the correlations are significant already on the time scales
shorter by an order of magnitude than in the case of the stock markets
(e.g.~\cite{lundin,reno}). In order to compare the size of correlations
between stocks of different transaction frequencies (being positively
correlated with market capitalization of the corresponding companies), we
select a few distinct sets of 30 stocks in such a way that the companies
within each set are characterized by similar capitalization (from $10^8$\$
to $10^{11}$\$). For each set of stocks we evaluate $\lambda_1(\Delta t)$
and compare it across the sets (Figure 3). The so-quantified correlation
range shows a systematic and monotoneous dependence on the capitalization:
for a given $\Delta t$, the larger the company, the stronger the average
coupling with its same-size counterparts. $\lambda_1(\Delta t)$ for the
DJI market being a mixture of $10^9 - 10^{11}$\$ companies is also
presented (diamonds in Fig.~3). In the case of the smallest firms
considered (triangles down), $\lambda_1$ is essentially at the RMT level
for all the time scales up to $\Delta t=30$ min. It is interesting to
note that a saturation level occurs only for the largest companies worth
at least $10^{11}$\$ each (circles) and for the DJI stocks; all the other
groups of companies display increase of $\lambda_1$ even for the largest
time scales analyzed. It may be hypothesized that the saturation could
manifest itself on time scales much longer than two days $-$ the smaller
the companies, the later the saturation. From this point of view it is
clear why $\lambda_1$ for the DJI stocks saturates earlier than for the
DAX stocks (Fig.2): as pointed out in~\cite{drozdz03b}, the average number
of transactions for the DAX companies is significantly smaller than for
the DJI ones. 

\begin{figure}
\epsfxsize 11cm
\hspace{1.0cm}
\epsffile{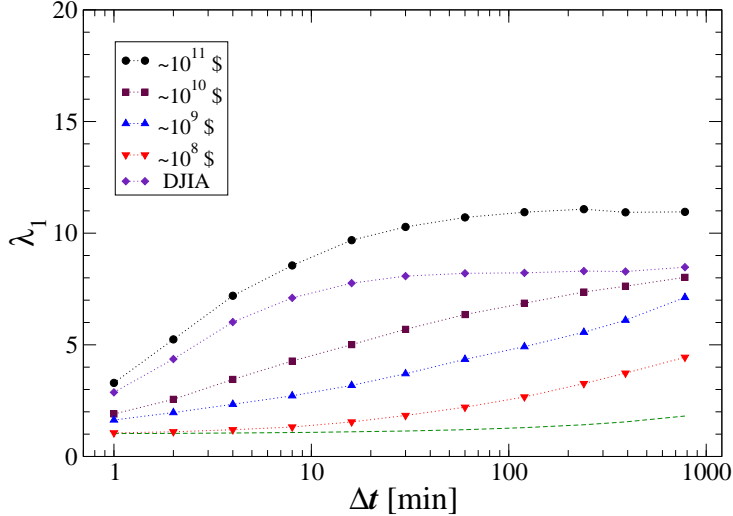}
\caption{Comparison of $\lambda_1(\Delta t)$ for several groups of 30
stocks representing companies of different capitalization. Noise level is 
indicated by dashed line.}
\label{fig:capital}
\end{figure}

One of the possible sources of the analyzed effect can be the lack of
synchronicity in transaction moments for different
securities~\cite{lo,reno,martens} and the associated nonsyncronicity of
their price determination. We choose a pair of the most correlated stocks
in DJIA: Citigroup (C) and General Electric (GE) and by following
ref.~\cite{reno} we remove from the original data all transactions which
didn't take place simultaneously for both stocks. (By ``simultaneous'' we
understand transactions which were made within the same second.)
Obviously, this strongly reduces the average number of transactions per
business day (e.g. for GE:  from 4240 to 440) but nevertheless there is
still more than one simultaneous transaction per minute. Next we proceed
in the usual way by creating a time series of $\Delta t$-returns and then
by calculating the correlation coefficient $C_{\rm C,GE}(\Delta t)$. 
Figure 4(a) displays the functional dependence of $C_{\rm C,GE}$ on the
time scale for the original data comprising all the transactions (circles)
and for the synchronous data only (squares). For $\Delta t < 60$ min,
values of the correlation coefficient for the synchronous data are
elevated in respect to the nonsynchronous data with the difference
increasing with decreasing $\Delta t$. For the shortest 1 min time scale,
the elimination of nonsynchronous transactions almost doubles $C_{\rm
C,GE}$. In Fig.~4(b) analogous calculation is carried out by using data
from the two most active NASDAQ stocks: Dell Computer (DELL) and Intel
Corp. (INTC). As the average nonsynchronous transaction number per a
business day exceeds 20,000 with over 10,000 simultaneous transactions per
day, here the correlations are detectable even at a-few-seconds time
scales. The two analyzed data sets differ from each other only for $\Delta
t < 30$ s; this difference is, however, impressive: for $\Delta t=1$ s
$C_{\rm DELL,INTC}$ is about an order of magnitude larger for the
synchronous data than in the other case. 

\begin{figure}
\epsfxsize 11cm
\hspace{1.0cm}
\epsffile{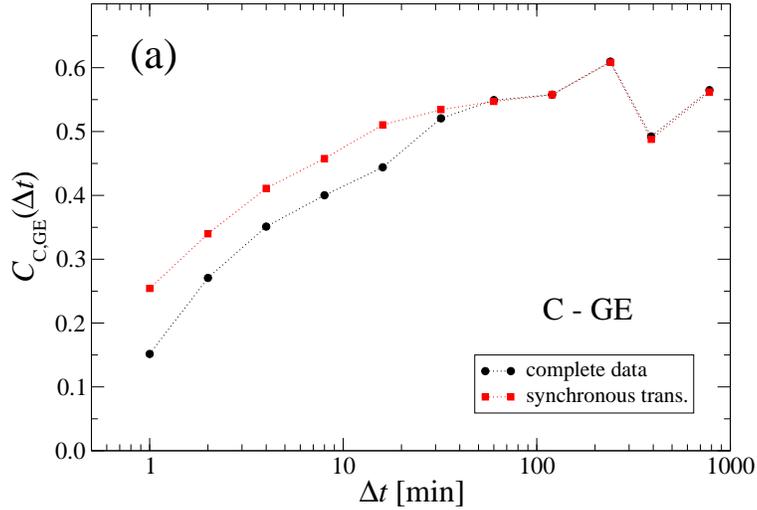}

\hspace{1.0cm}
\epsfxsize 11cm
\epsffile{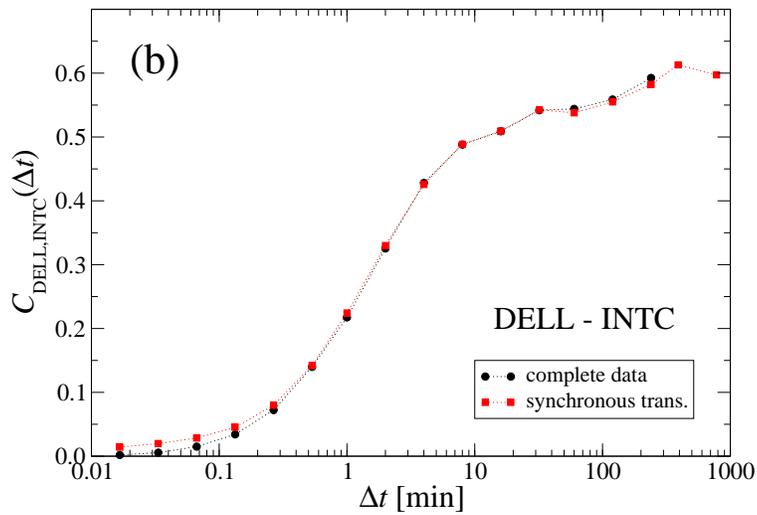}
\caption{Correlation coefficient $C_{i,j}$ as a function of the time scale
$\Delta t$ for a pair of moderately active DJI stocks (a) and for a pair
of the most frequently traded NASDAQ stocks (b). In both (a) and (b)
results for the following two data sets are illustrated: the complete
signals comprising all transactions (circles) and the modified signals
representing simultaneous transactions only (squares).}
\label{fig:frequency}
\end{figure}

A straightforward generalization of this procedure towards more than two
degrees of freedom (stocks) requires some care, though. Selection of
precisely synchronous transactions leads to a drastic reduction of the
time resolution and of the range of available time scales, therefore this
proves inefficient for more than two stocks. In order to overcome this
problem, we weaken our definition of synchronicity by introducing a
tolerance parameter $\tau$; now transactions are considered to be
synchronous if they are made within an interval
$(t_{\alpha}-\tau,t_{\alpha}+\tau)$, where $t_{\alpha}$ stands for the
transaction time for a reference stock $\alpha$ being the most active
stock in a set of the stocks under study. The functional dependence of
$\lambda_1$ on $\Delta t$ for a set of the 10 most frequently traded
stocks and for several values of $\tau>0$ is presented in Figure 5. It is
clear from the figure that the more synchronous transactions are
considered ($\tau \rightarrow 0$), the stronger short-time-scale couplings
among the stocks occur. 

\begin{figure}
\epsfxsize 11cm
\hspace{1.0cm}
\epsffile{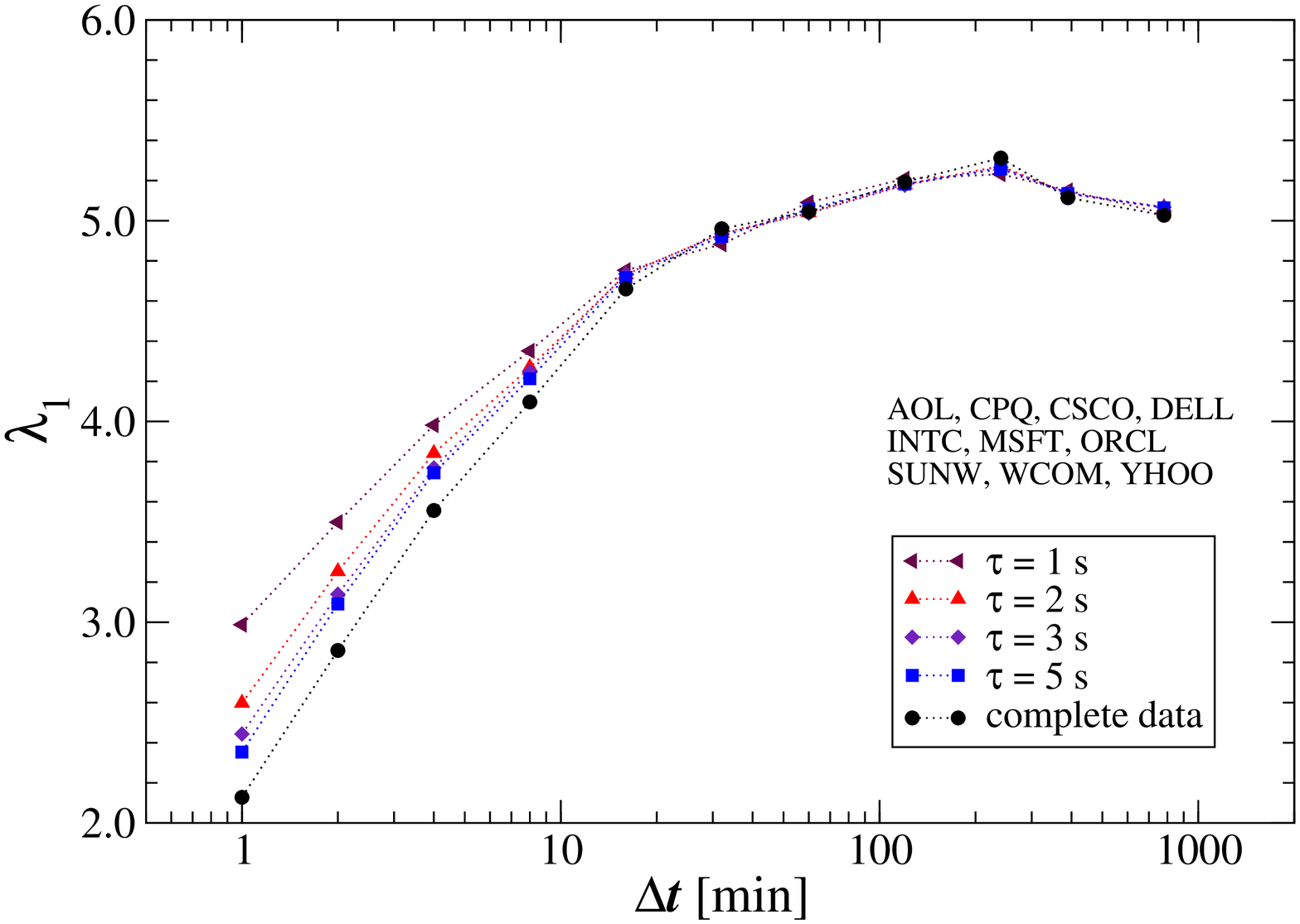}
\caption{Behaviour of $\lambda_1(\Delta t)$ for different values of
tolerance $\tau$ (see text for explanation) for a group of the 10 most
active stocks.}
\label{fig:mostfreq}
\end{figure}

Nonsynchronicity of trading being related to the microscopic randomness of
the stock market dynamics cannot alone account for the observed size of
the correlations decay and thus there must be other influential factors
here, like for example the possible existence of lagged correlations
amongst the assets~\cite{epps,kullmann,reno,martens} or the differences in
the time horizon of trading strategies of individual market
agents~\cite{mueller}. Since these factors cannot be directly implemented
in the correlation matrix formalism, we shall not discuss this issue in
the present paper and instead we refer reader to the literature (see also
e.g.~\cite{lo,lundin}).

\begin{figure}
\epsfxsize 11cm
\hspace{1.0cm}
\epsffile{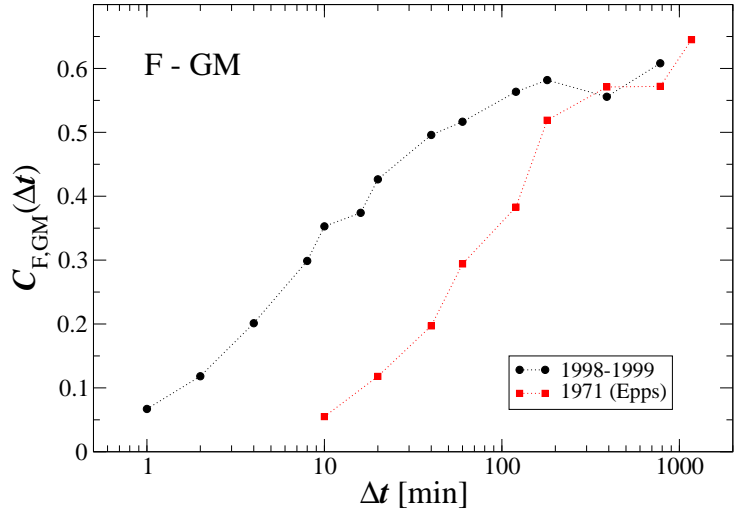}
\caption{Correlation coefficient for F and GM as a function of time
scale. Two data sets are presented: 1998-1999 denoted by circles, and 1971
(after Epps~\cite{epps}) denoted by squares.}
\label{fig:epps}
\end{figure}

Owing to the fact that the high frequency data from the American stock
market has been extensively studied over past decades, we can compare
outcomes from our study with the ones from other studies. In his original
paper~\cite{epps}, Epps investigated data for the stocks of the automobile
sector, recorded during a few months of 1971. In Figure 6 we confront the
Epps' historically distant results with the more contemporary ones from
1998-1999 for an exemplary pair of stocks (Ford Motor - F, General Motors
- GM), studied also in ref.~\cite{epps}. The phenomenon of a decrease of
the correlation coefficient with decreasing $\Delta t$ is evident in both
cases, but at present it is strongly shifted towards the shorter time
scales. In contrast, for the daily time scales $C_{\rm F,GM}$ assumes
comparable magnitudes for both 1971 and 1998-99. In another study Andersen
et al. (ref.~\cite{andersen}), who analyzed data from the time period
1993-1998, showed that the average value of the correlation coefficient
calculated for 30 DJI stocks assumes 0.12 at the time scale of 5 min. In
our case, this is equivalent to the averaged correlation matrix element,
which for $\Delta t=5$ min equals 0.19. Although this comparison is not
fully decisive because of only one time scale investigated, it suggests
that the short-time-scale couplings among the DJI stocks were stronger at
the end of 1990's than they used to be on average during this decade;
again, this is in the spirit of the above conclusions.

\section{Conclusions}

For a summary, we study the process of the emergence of a collective
market out of noise by investigating a time-scale dependence of the
magnitude of the inter-stock correlations for the companies listed in the
American and German markets. We observe that the correlations' magnitude
quantified in terms of the Pearson's correlation coefficient increases
with increasing the time scale from a statistically insignificant level
at the time scales comparable with the average inter-transaction interval
to high values at the long hourly or daily time scales. By applying the
correlation matrix formalism to the high-frequency data, in a simple way
we generalize these results to more than two degrees of freedom. We show
that such behaviour of the inter-stock correlations can be observed also
globally for the market as a whole. Our results convince us that one of
the most important factors that determines the time scales at which the
collective behaviour of assets occur is the trading frequency: for a given
time scale, the most active stocks are also among the most correlated
ones. In contrast, the stocks of small companies which are characterized
by a low number of transactions, present non-significant correlations up
to daily time scales. We next demonstrate that the synchronous data with a
supressed level of randomness of transaction moments reveals stronger
couplings than the original data. Finally, our results provide us with the
indication that nowadays the collective market emerges at significantly
shorter time scales than it used to do in the more distant history, i.e.
for the large American companies the market shows the trace of a weak
collectivity already at the minute or even the intra-minute time scales
compared with the intra-hour scales previously.

This result recalls a congenial effect of a faster convergence of the
stock returns distributions towards a Gaussian in the contemporary data if
compared with the historical data. As documented in ref.~\cite{drozdz03b},
in the same period 1998-99, the scaling law with the exponent $\alpha
\simeq 3.0$ breaks already at the time scales of tens of minutes, while
earlier it was still hold even at the time scales of several
days~\cite{plerou99b}. We interpreted this phenomenon as being a direct
consequence of a faster information processing and a faster loss of memory
in the evolution of the market when going from past to present. Now, while
looking at correlations among the stocks at various time scales, we
receive a further firm support for such conclusions. Due to the fact that
nowadays the emergence of the collective dynamics of stocks can be
observed at time scales much shorter then before, and keeping in mind that
one of its fundamental governing factors is an asset's trading frequency,
we shall underline that a time flow in the stock market (and possibly in
other financial markets as well) is not constant over long periods but
instead, owing to the technological progress and an enlarged flow of the
arriving information, the market time tends to accelerate: effectively,
one day in 1980 might not be completely equivalent to one day in 2000. A
straightforward and far-reaching consequence of this fact is that
potential models of the financial dynamics which do not take this
acceleration into account and which assume that properties of the
market dynamics are time-invariant might be not fully adequate. It is also
interesting to notice that such an acceleration of the financial market
evolution as viewed from the linear time scale is consistent, at least
qualitatively, with a scenario priovided by the log-periodicity effect,
especially the one that refers to the last 200 years~\cite{drozdz03a}.

\end{document}